\documentclass[conference]{IEEEtran}
\IEEEoverridecommandlockouts
\usepackage{placeins}
\usepackage{cite}
\usepackage{amsmath,amssymb,amsfonts}
\usepackage{algorithmic}
\usepackage{graphicx}
\usepackage{textcomp}
\usepackage{xcolor}
\usepackage{dblfloatfix}    % To enable figures at the bottom of page
\usepackage{listings}
\usepackage[normalem]{ulem}
\usepackage{tabularx}
\usepackage{xcolor,colortbl}
\usepackage{array}
\definecolor{codegreen}{rgb}{0.6,0.6,0.6}
\definecolor{codegray}{rgb}{0.5,0.5,0.5}
\definecolor{codepurple}{rgb}{0.58,0,0.82}
\definecolor{backcolour}{rgb}{0.95,0.95,0.95}
\lstdefinestyle{mystyle}{
    backgroundcolor=\color{backcolour},   
    commentstyle=\color{codegreen},
    keywordstyle=\color{blue},
    numberstyle=\tiny\color{codegray},
    stringstyle=\color{codepurple},
    basicstyle=\ttfamily\footnotesize,
    breakatwhitespace=false,         
    breaklines=true,                 
    captionpos=b,                    
    keepspaces=true,                 
    numbers=left,                    
    numbersep=5pt,                  
    showspaces=false,                
    showstringspaces=false,
    showtabs=false,                  
    tabsize=2
}

\def\BibTeX{{\rm B\kern-.05em{\sc i\kern-.025em b}\kern-.08em
    T\kern-.1667em\lower.7ex\hbox{E}\kern-.125emX}}
\begin{document}

\title{Enabling Efficient Hardware Acceleration of Hybrid Vision Transformer (ViT) Networks at the Edge}

\author{
\IEEEauthorblockN{{Joren Dumoulin, Pouya Houshmand, Vikram Jain and Marian Verhelst}}
\IEEEauthorblockA{{MICAS, ESAT, KU Leuven}}
\IEEEauthorblockA{Email: joren.dumoulin@kuleuven.be}
}

\maketitle

\begin{abstract}
Hybrid 
vision transformers combine the elements of conventional neural networks (NN) and vision transformers (ViT) to enable lightweight and accurate detection. 
However, several challenges remain for their efficient deployment on resource-constrained edge devices.  
The hybrid models suffer from a widely diverse set of NN layer types and large intermediate data tensors, hampering efficient hardware acceleration. To enable their execution at the edge, this paper proposes innovations across the hardware-scheduling stack: a.) At the lowest level, a configurable PE array supports all hybrid ViT layer types; b.) temporal loop re-ordering within one layer, enabling hardware support for normalization and softmax layers, minimizing on-chip data transfers; c.) further scheduling optimization 
employs layer fusion across inverted bottleneck layers to drastically reduce off-chip memory transfers. 
The resulting accelerator is implemented in 28nm CMOS, achieving a peak energy efficiency of 1.39 TOPS/W at 25.6 GMACs/s.

\end{abstract}

\section{Introduction}

\IEEEPARstart{N}{eural} network models for computer vision have experienced significant advancements in terms of accuracy, size, and complexity \cite{convnext, vit, resnet, vggnet}. Striving for ever-increasing accuracy scores on increasingly complex tasks, 
the number of parameters in these models can easily reach billions, making them impractical to store and use on edge devices. 
Consequently, there has been a parallel evolution focusing 
on improving the energy efficiency and throughput of the models \cite{mobilenet, mobilenetv2, mnasnet}. These compact models require relatively fewer parameters ($<$5M) and computations ($<$2G), resulting in faster and more efficient inference.  
While these evolutions were initially mainly exploring 
traditional convolutional neural networks (CNNs) \cite{resnet, mobilenetv2}, vision transformers (ViT) \cite{vit} are proposed as a promising alternative to CNNs for high-performance applications. Although the original ViT models do not scale down well to edge execution scenarios \cite{mobilevit, deit}, a more efficient breed of ViT models is recently emerging. Hybrid ViTs \cite{mobilevit, mobilevitv2, edgenext} combine the properties of vision transformers and convolutional neural networks, to achieve impressive accuracies at very low model sizes, outperforming CNN networks \cite{mobilevit}.

\begin{figure}[h]
    \centering
    \includegraphics[width=\linewidth]{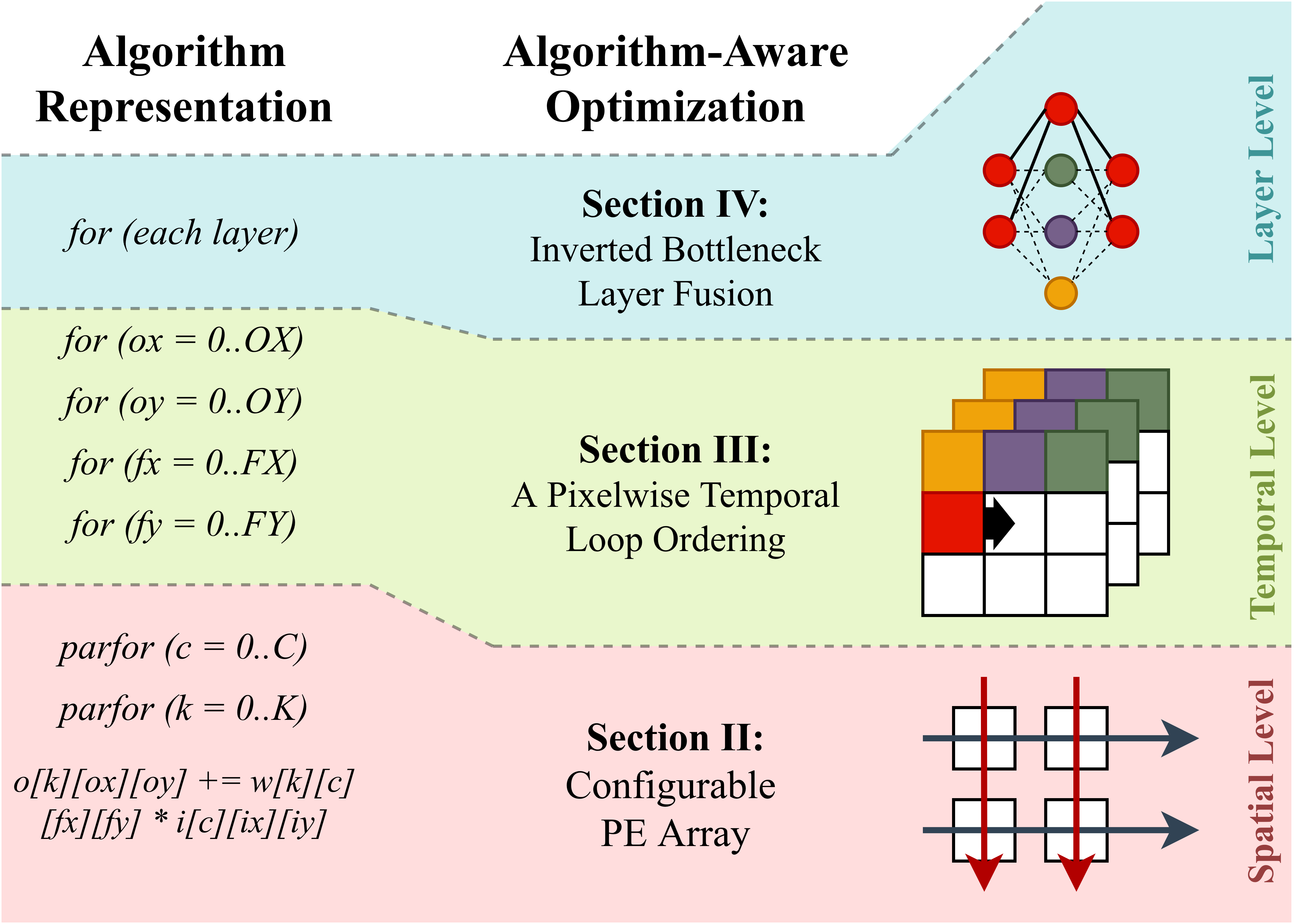}
    \caption{For loop representation of the algorithm and the corresponding optimizations leading to efficient hardware execution of the network workload. Parfor indicates the parallel execution of the for loops.}
    \label{fig:titlefig}
\end{figure}

Besides the neural network model, the hardware platform can be also optimized for efficient inference at the edge. As conventional CPUs are not suited to efficiently execute neural network workloads \cite{horowitz, hameed}, customized processors are developed that exploit the deterministic dataflow and widely parallel nature of the computations 
in CNN inference 
\cite{envision, diana, tinyvers, vaqf, rwa, vitcod, moh, fick, yuhao, syntiant, unpu}. 
However, so far no work has been published that addresses the development of architectures for hybrid networks, whose hardware acceleration potential is yet to be uncovered.

\begin{figure}[b]
    \centering
    \includegraphics[width=\linewidth]{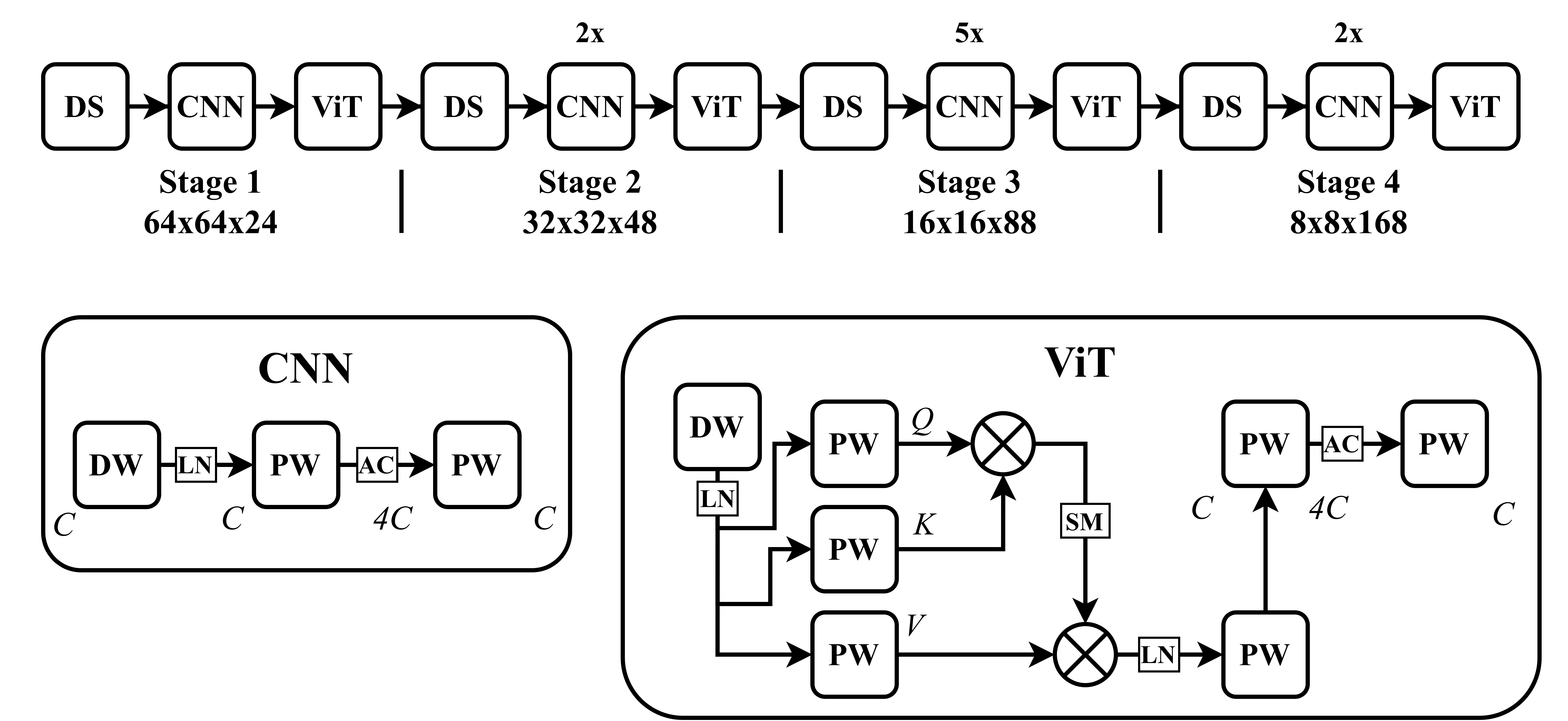}
    \caption{Diagram of the EdgeNeXt \cite{edgenext} network, consisting of downsampling layers (DS), convolutional blocks (CNN), and vision transformer blocks (ViT). Pointwise convolutions are denoted as PW, depthwise convolutions as DW. Downsampling Layers (DS) are implemented using strided convolutions. Nonlinear functions are LN (LayerNorm), AC (Activation), and SM (Softmax).}
    \label{fig:edgenext}
\end{figure}

\begin{figure*}[b!]
    \centering
    \includegraphics[width=\textwidth]{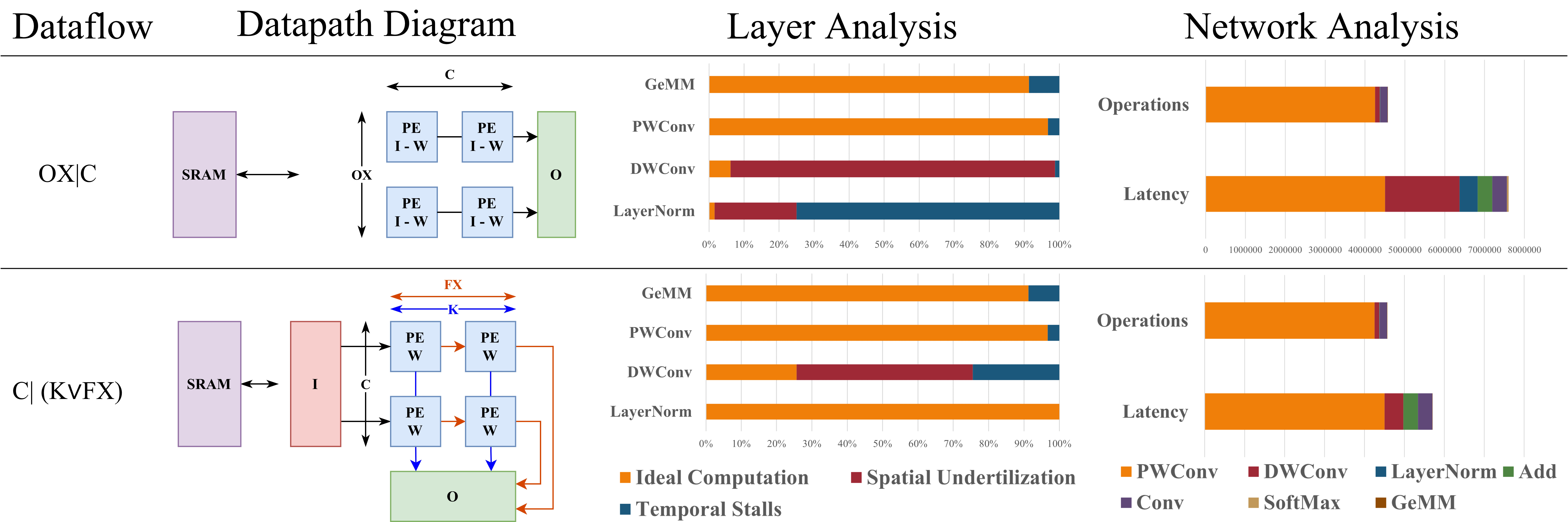}
    \caption{Results of the design space exploration experiments. The architectures considered use a PE array with dimensions 16$\times$16. The layer analysis shows the lost cycles to spatial underutilization and temporal stalls, in comparison to the actual executed computations.  
    The network analysis shows the distribution of the number of operations in the network, compared to the accelerator latency of the entire network.}
    \label{fig:concepts}
\end{figure*}

The target hybrid model used for benchmarking is EdgeNext-S \cite{edgenext}, which is representative for a wide class of edge hybrid ViTs and shown in Fig.~\ref{fig:edgenext}. The model exhibits a wide variety of layer types - convolutional, matmul, depthwise, pointwise, elementwise - requiring a reconfigurable acceleration architecture.

In this work we present a hardware accelerator for hybrid ViT networks, innovating at three different levels of abstraction, summarized in Fig. \ref{fig:titlefig}. Section~\ref{dse_spatial} introduces a reconfigurable spatial mapping  to maximize data reuse between processing elements, regardless of layer types. A temporal loop optimization for memory efficiency is presented in section~\ref{dse_temporal}, while

section~\ref{layerfusion} further optimizes the execution schedule to minimize external data transfers for inverted bottleneck layers. 

 Finally, in section~\ref{implementation}, we present and benchmark a, parametrizable, programmable implementation of the accelerator.

\section{Reconfigurable processing array for hybrid model acceleration}
\label{dse_spatial}

The computation of each neural network layer can be represented by a set of nested for-loops, as illustrated in Fig.~\ref{fig:titlefig}. This representation allows to express the parallelization dimensions, that are spatially unrolled across a computational array. For a 2D array, the \textit{dataflow} (Spatial Unrolling~X)$\vert$(Spatial Unrolling~Y) specifies the for-loops mapped onto the array. 
The remaining nested loops that are not parallelized are executed sequentially in time. The \textit{temporal loop mapping} defines how these loops are \textit{tiled} and \textit{reordered} \cite{zigzag}, and how the input, weight and output operands per layer are distributed across the different caching levels in the memory hierarchy. 
A well-designed mapping combines the optimal spatial mapping with its optimal temporal ordering, jointly maximizing the spatial and temporal reuse of input, weight, and output operands, thereby reducing memory access costs across the memory hierarchy.

To uncover the most energy- and latency-efficient mapping across the layers of hybrid ViTs we apply the design space exploration tool ZigZag \cite{zigzag} on a high-level description of a DNN accelerator hardware template. 
Fig.~\ref{fig:concepts} summarizes the results of the design space exploration on two different spatial dataflow architectures. For each architecture, the performance at the layer-level (across different layer types) and at the full hybrid ViT network level is analysed. The top architecture is a standard, single dataflow accelerator implementation, computing all the layers using a fixed spatial mapping ($OX \vert C$). While this architecture is efficient for the common convolutional layers and matrix multiplications, it shows very poor (spatial and temporal) array utilization for depthwise convolutional layers. Even though the contribution of depthwise convolutional layers is minimal when considering all network layers, the poor mapping results in a very large latency overhead.

To solve this issue, we explore a secondary dataflow, supported by a reconfigurable array that can alternate between spatial mappings $C\vert K$ and $C\vert FX$ (denoted as $C\vert \left( K \vee FX \right)$), depending on the layer type.
This alternative dataflow drastically enhances support for depthwise kernels, leading to $18\%$ savings in latency compared to the fixed dataflow design.

 \section{Pixelwise Temporal Loop Ordering}
 \label{dse_temporal}
Beyond the optimization of the spatial dataflow, it is also necessary to improve the temporal loop ordering for maximal data stationarity. Additionally, we also want to include optimization of non-linear functions in the network, such as softmax and LayerNorm. As can be seen in Fig.~\ref{fig:concepts} (top), the layernorm contributes to a significant latency overhead, whilst the number of operations of this layer is negligible.

These layers typically require some dynamic calculation of layer statistics along the C dimension (Eqn. (\ref{eq:layernorm})).

\begin{equation}
    \label{eq:layernorm}
    E[X] = \sum_{c=0}^C{X[ox][oy][c]} \qquad    \forall ox, oy
\end{equation}
 This requires streaming the complete data in and out of the SRAM cache, resulting in latency and energy overhead. Instead, we choose a temporal loop ordering which allows us to compute the layernorm along with the computational layer that precedes it, fusing the two layers into a single computation flow. For this, we employ a pixelwise temporal loop ordering for the computational layer, which processes the data pixel by pixel, all channels at the same time.
The pixelwise loop ordering, in combination with the spatial mapping of the C and K dimensions, makes sure that the order of outputs follows the loop ordering as shown in Listing \ref{listingcnn}.
\begin{lstlisting}[language=C, caption=Order of outputs using a pixelwise temporal loop ordering., label=listingcnn, style=mystyle]
for (x = 1 .. X)         //for each column
  for (y = 1 .. Y)       //for each row
     for (c = 1 .. C)    //for each channel
\end{lstlisting}
Before writing back the output data to local storage, the data can then be buffered, to allow for direct implementation of the non-linear layers. With the selected pixelwise temporal loop ordering, only the innermost for-loop must be buffered, to contain all the information required for the computation of layer statistics. This parallel implementation of the non-linear layers can remove their latency and energy overhead on system performance.

\section{Inverted Bottleneck Layer Fusion}
\label{layerfusion}
\begin{figure}
    \centering
    \includegraphics[width=\linewidth]{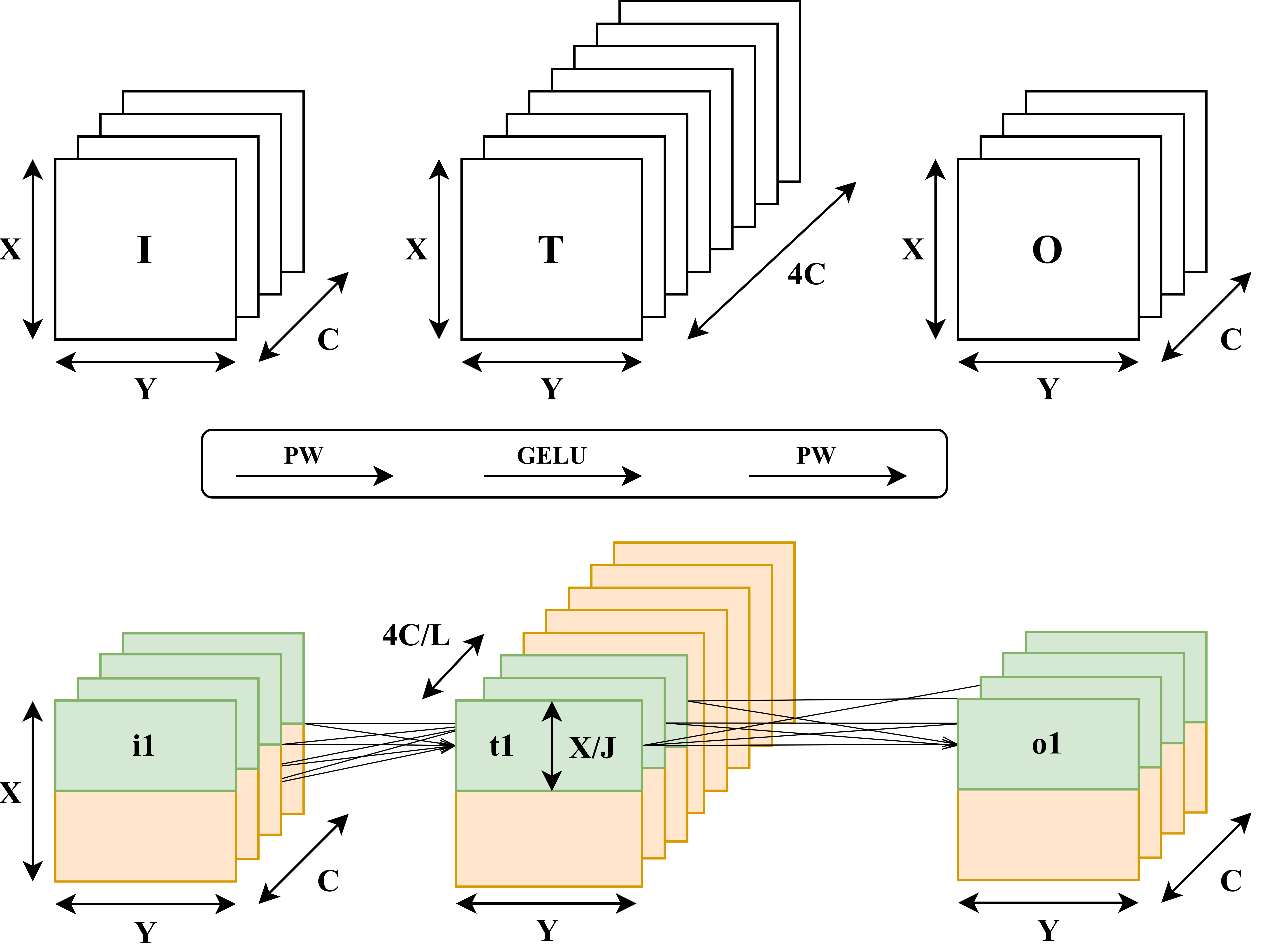}
    \caption{The inverted bottleneck structure (top) and a visualization of the layer fusion mechanism (bottom).}
    \label{fig:layerfusion}
    \vspace{-0.3cm}

\end{figure}

\begin{figure}
    \centering
    \includegraphics[width=\linewidth]{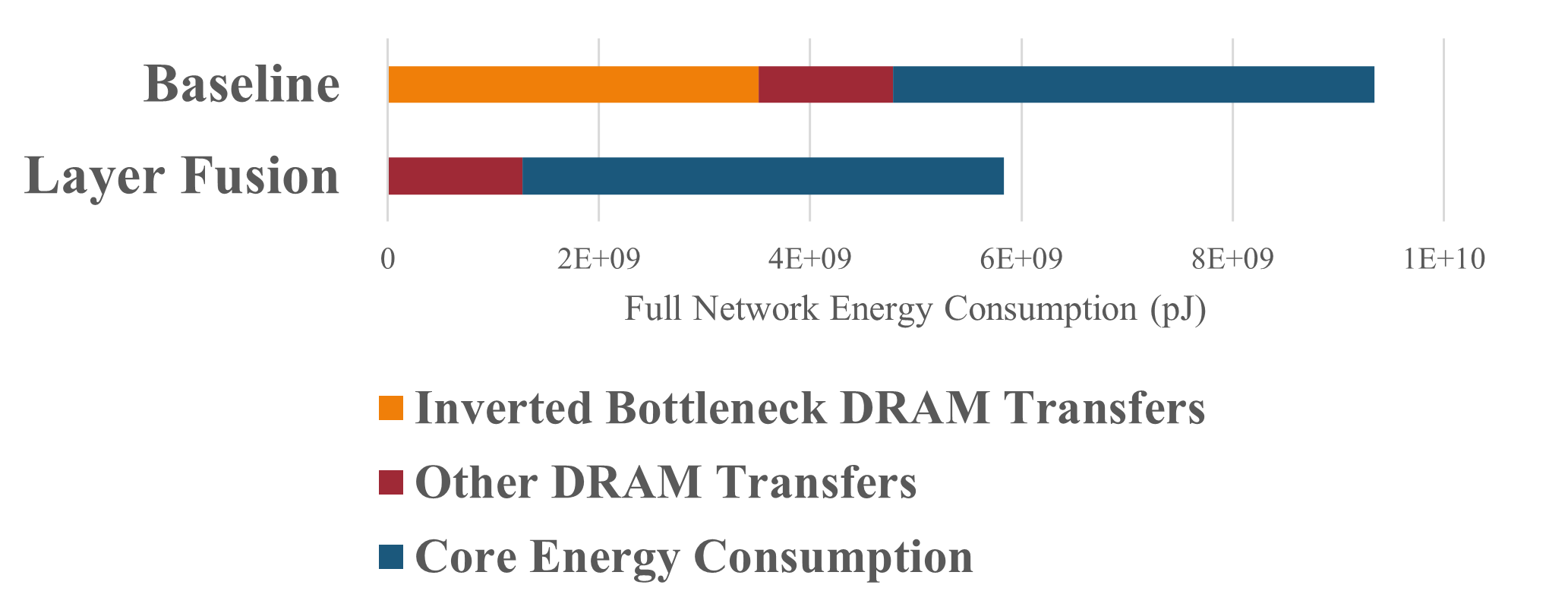}
    \caption{Estimated layer fusion gains using a DRAM access cost of 100 pJ/byte. The inverted bottleneck structure accounts for 63.6\% of all DRAM transfers in the EdgeNeXt-S network. With the addition of the layer fusion mechanism total system energy consumption reduces with 37.6\%.}
    \label{fig:layerfusiongain}
        \vspace{-0.3cm}

\end{figure}

With the two previous innovations drastically reducing the spatial stalls and the stalls due to on-chip memory access, we are left with remaining overheads from accessing off-chip DRAM. Especially the inverted bottleneck structure present in many neural networks \cite{mobilenetv2, convnext,edgenext, mobilevit, vit}, gives rise to excessive DRAM communication. This structure, shown in Fig.~\ref{fig:layerfusion}, realizes a temporary expansion in the channel count (usually a factor of 4), that results in a large intermediate activation map. Due to on-chip memory resource constraints, these intermediate activations typically exceed on-chip SRAM capacity, necessitating DRAM memory transfers with an energy cost per bit being orders of magnitude above SRAM transfers \cite{horowitz}.
With an estimated access cost of 100 pJ/byte, ZigZag estimates the DRAM memory costs for the EdgeNeXt-S model accounting for up to 52\% of total energy usage (Fig.~\ref{fig:layerfusiongain}).

To mitigate this, we propose a layer fusion method that lifts the need to store intermediate activations in the inverted bottleneck structure, thereby drastically enhancing system energy efficiency.
The proposed mechanism, shown in Fig.~\ref{fig:layerfusion} (bottom), targets depth-first execution of the two constituting layers of the subnetwork -- two stacked pointwise convolutions with an activation function in between -- without storing the entire intermediate feature map. This is achieved by tiling the intermediate feature map $T$ along the $X$ and $C$ dimension, and repeatedly performing following two steps: 
\begin{enumerate}
    \item As soon as a tile $t1$ of the intermediate feature map $T$ becomes available, it is stored in local memory.
    \item The tile $t1$ is immediately reused towards the computation of partial results of a tile $o1$ of the output feature map $O$. After accumulating these results to the $o1$ output values, the intermediate tile $t1$ can be discarded to make room in the local buffer.
\end{enumerate}

The tile sizes are optimized using ZigZag, resulting in the energy benefits of the layer fusion mechanism on the entire network workload shown in Fig.~\ref{fig:layerfusiongain}.

\newpage

\section{Open-source Parametrized \\ Hardware Implementation}
\label{implementation}

\begin{figure}[t]
    \centering
    \includegraphics[width=\linewidth]{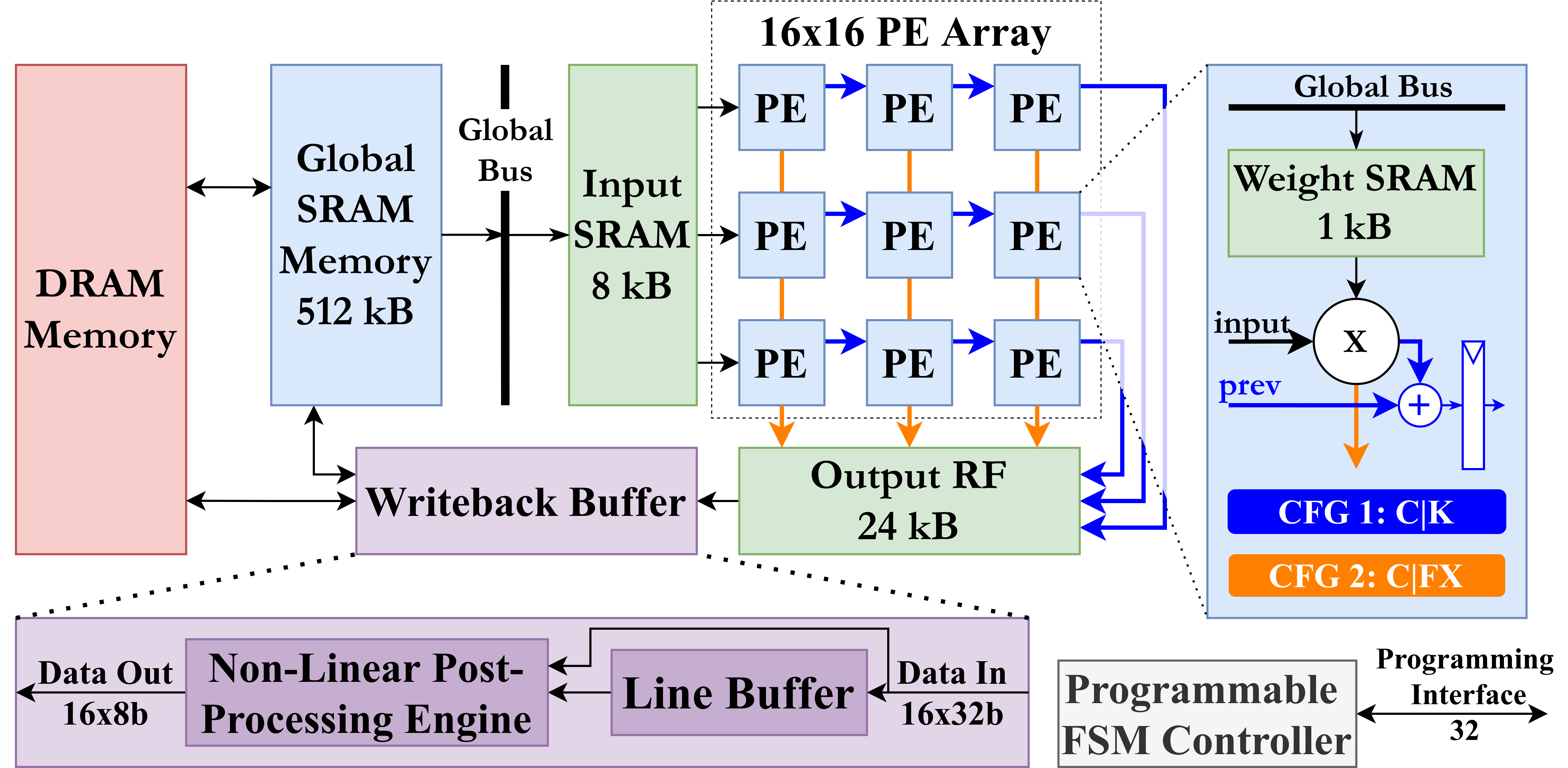}
    \caption{Hardware schematic of the proposed neural accelerator.}
    \label{fig:hardwareschematic}
\end{figure}

The proposed innovations are integrated into a neural accelerator implementation. Fig.~\ref{fig:hardwareschematic} depicts the overall architecture, centered around a $16\times16$ PE Array, working on 8-bit data. Every PE element has a local weight memory to allow for unicast access of the weights. A local 8 kB input memory multicasts activations along a single dimension of the PE array to support the derived $C\vert \left( K \vee FX \right)$ dataflow. The 32-bit accumulation outputs are sent to a 24 kB output register file. A global 512 kB on-chip SRAM memory is used by all of the lower memory levels in a user-programmable way. A 128-bit data bus acts as the data interface to an off-chip DRAM memory. The PE array dimensions, bus widths, memory sizes, and depthwise layer support are all parameterized, and the accelerator is functionally tested on different configurations. The implementation is available at https://github.com/KULeuven-MICAS/hyvit-accelerator.

\subsection{Reconfigurable processing array}
The datapath supports the two spatial dataflows introduced in Section \ref{dse_spatial}. 
The first configuration realizes the $C\vert  K$ flow, targeting regular/pointwise convolutions and GeMM operations by accumulating outputs along the PE array columns using adder trees. The second configuration targets depthwise convolutions, implementing a $C\vert FX$ spatial mapping, where the output data propagates along the rows of the PE array, and results are accumulated across clock cycles. 
Supporting both dataflows has no impact on the existing memory hierarchy, resulting in only minimal area overhead. 

\subsection{Pixelwise temporal loop support}
The writeback buffer serves two main purposes. Firstly, it handles bus contentions with the DRAM and global SRAM memories, writing back data to memory when this bus is idle. Secondly, it allows for the implementation of a non-linear post-processing engine. Using the pixelwise temporal dataflow proposed in Section \ref{dse_temporal}, the line buffers are large enough to contain all the information needed to effectively implement a wide variety of layers such as LayerNorm and SoftMax, which are widely used in vision transformers. It can also handle data quantization and activation functions.

\subsection{Layer fusion support}
The datapath and memory units are controlled by a central FSM controller which is highly instruction programmable by a 32-bit instruction bus. The programmability allows for a very wide variety of layer sizes. The source and target location of all layer operands are programmable across the entire memory hierarchy. The layer fusion mechanism proposed in Section \ref{layerfusion} is as such supported by the programmability of this controller. 

\subsection{Synthesis Results}

\begin{figure}
    \centering
    \includegraphics[width=.8\linewidth]{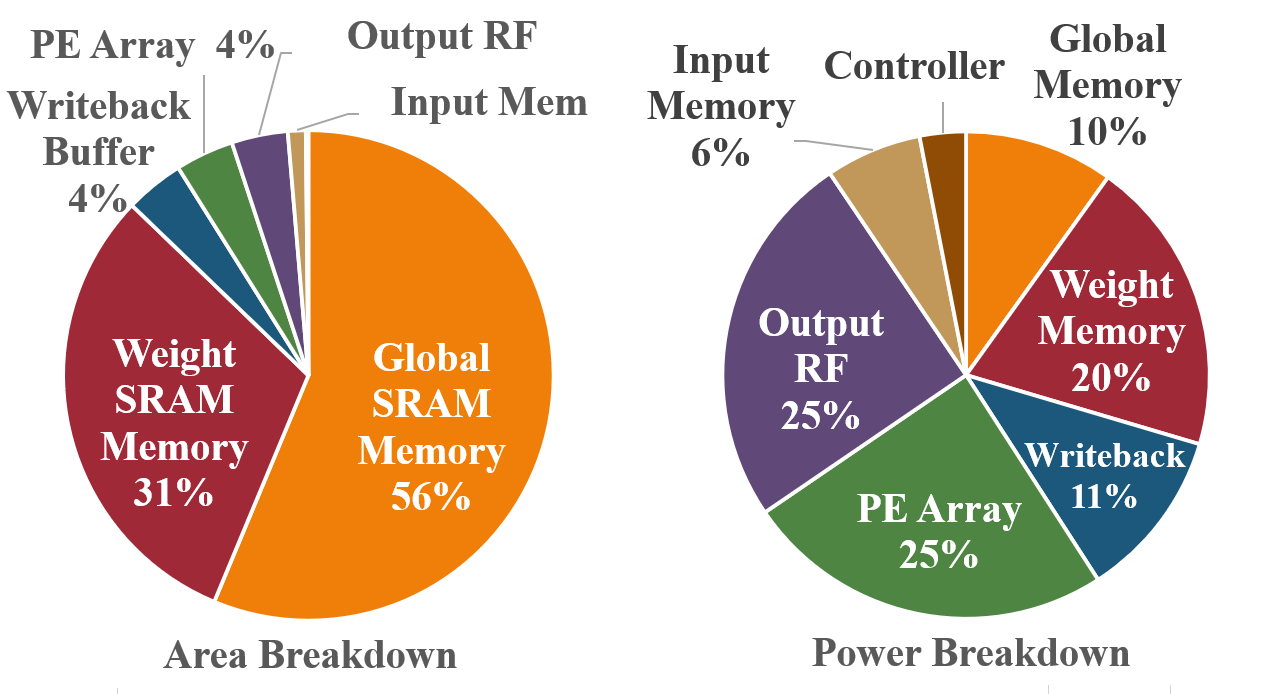}
    \caption{Area breakdown (left) and Power breakdown for the computation of a pointwise convolutional layer, at a clock frequency of 100MHz and a peak throughput of 25.6GMACS/s (right).} 
    \label{fig:results}
    \vspace{-0.2cm}
\end{figure}

The design is synthesized in TSMC 28HPC+, for area estimations and power simulations. 
The synthesised design results in a chip area of $1.48 mm^2$, and a full area breakdown of the accelerator is shown in Fig.~\ref{fig:results}. The hardware overhead of the configurable array, at just 1.1\%, is included as part of the PE array. 

The chip runs at a nominal clock frequency of 100 MHz, achieving a peak energy efficiency of 1.39 TOPS/W, at a throughput of 25.6 GMACs/s. A power breakdown of the accelerator is shown in Fig.~\ref{fig:results}.

\begin{figure}
    \centering
    \includegraphics[width=.8\linewidth]{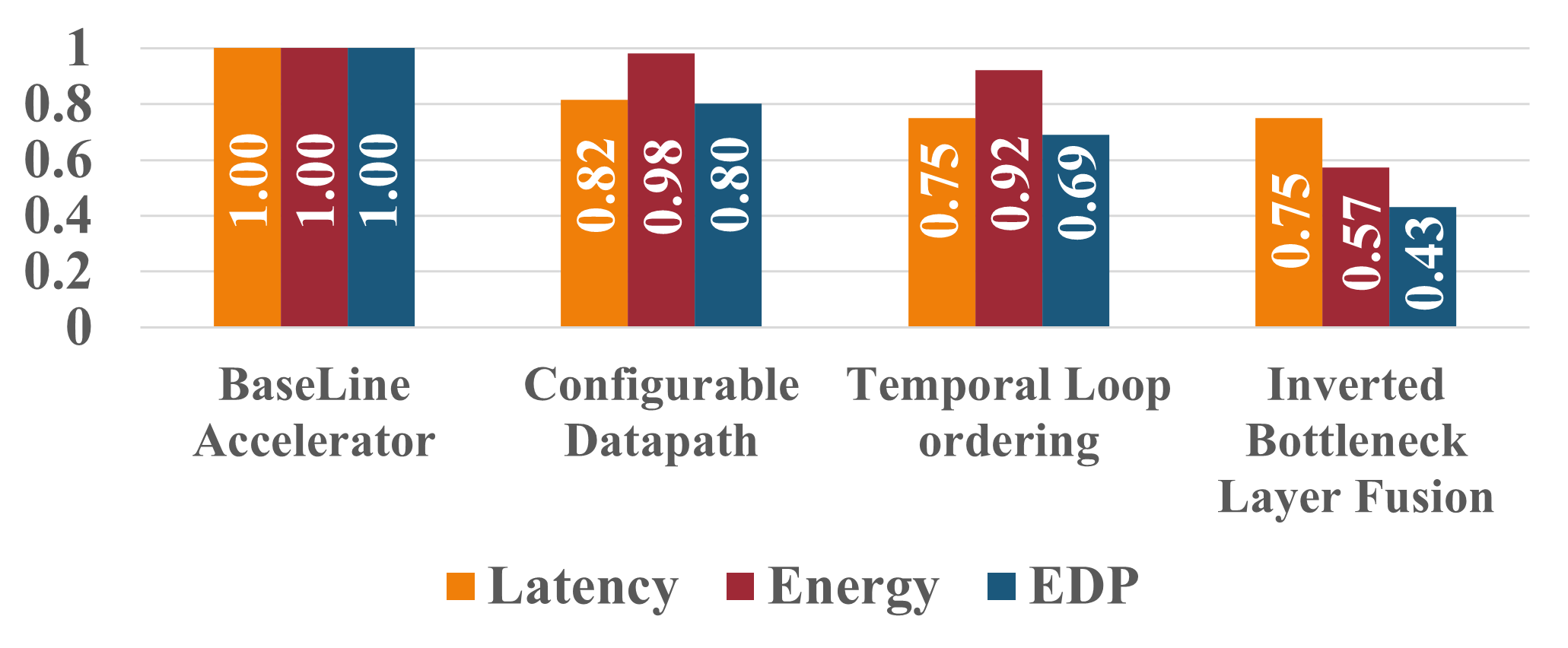}
    \caption{Normalized contributions of the proposed optimizations on system latency, energy, and EDP for a full network workload.}
    \label{fig:contribution-results}
    \vspace{-0.3cm}
\end{figure}

The main contributions of this work are targeted towards full network optimization of hybrid vision transformers. The final improvements over the baseline accelerator from section \ref{dse_spatial} are summarised in Fig.~\ref{fig:contribution-results}. The accelerator is able to run the target network EdgeNeXt-S at 13.16 FPS, at a power consumption of just 18.4 mW. This results in an impressive network efficiency of 731.1 FPS/W. Table \ref{table:fullnetwork} compares the accelerator to other works, further emphasizing the flexibility and full-network efficiency of this implementation.

\definecolor{tablegray}{gray}{0.9}

\begin{table}
    \vspace{-2mm}
    \centering
    \scriptsize
    \caption{Comparison to other works. 
    }
    \begin{tabular}{>{\centering\arraybackslash}m{1.6cm} | >{\centering\arraybackslash}m{0.7cm} >{\centering\arraybackslash}m{0.7cm} >{\centering\arraybackslash}m{0.7cm} >{\centering\arraybackslash}m{0.7cm} >{\centering\arraybackslash\columncolor{tablegray}}m{0.8cm} >{\centering\arraybackslash}m{0.7cm}}
               & \multicolumn{4}{c}{CNN Accel.} & { Hybrid \newline Accel.} & ViT\newline Accel. \\ \hline
               & \cite{unpu} \newline \tiny JSSC'19         & \cite{envision} \newline \tiny ISSCC'17 & \cite{moh} \newline \tiny ISSCC'21 & \cite{syntiant} \newline \tiny ISSCC'23   & This Work           &     \cite{vaqf}             \\ \hline
    \textbf{Technology} & 65 nm         & 28 nm & 28nm & 40nm &  28 nm    (synth.)        & 16 nm FPGA       \\ \hline
    \textbf{Bit Precision} & 1-16 & 1-16 & 8 & 8-16 &  8 & 6-32 \\ \hline
    \textbf{Network Efficiency {(FPS/W)} @Accuracy $^{\mathrm{a}}$} & 61.6 @71.5\% & 64.2 @71.5\% & 323.2 @76.9\% & 6024 @40.1\% &  \textbf{731.1} @78.8\% &  2.85 @81.5\%  \\ \hline
    \textbf{Power Cons. (mW)} & 3.2 - 297 & 7.6-290 & 125.8 & 0.83 &  18.4 & 8700 \\ \hline
    \multicolumn{6}{l}{\textbf{Layer Support}} \\
    \hline
    \textbf{GeMM \& Conv} & \checkmark & \checkmark & \checkmark & \checkmark & \checkmark &  \\ \hline
    \textbf{DWConv} & & & \checkmark & \checkmark  & \checkmark & \\ \hline
    \textbf{ViT} & & &  & & \checkmark & \checkmark \\ \hline
    \multicolumn{7}{l}{ \tiny  $^{\mathrm{a}}$Imagenet Top-1 Accuracy of targeted network.}

    \end{tabular}
    \label{table:fullnetwork}
\end{table}

\section{Conclusion}
This work presented three hardware-scheduling co-optimizations and a supporting hardware implementation for the acceleration of hybrid vision transformer networks. By addressing challenges related to the diverse layer types, and costly memory accesses, the proposed architecture demonstrates significant energy and latency improvements at a system level. 
Implementation of the proposed architecture in TSMC 28HPC+, achieves a peak efficiency of 1.39 TOPs/W.

\newpage

\section*{Acknowledgements}

This project has been partly funded by the European Research Council (ERC) under grant agreement No. 101088865, the European Union’s Horizon 2020 programme under grant agreement No. 101070374, the Flanders AI Research Program and KU Leuven.
 
\bibliographystyle{IEEEtran}
\bibliography{bibtex}

\end{document}